\documentclass[12pt,preprint]{emulateapj}

\shorttitle{Origin of molecular oxygen in Comet 67P/Churyumov-Gerasimenko}
\shortauthors{Mousis et al.}

\usepackage{etex}
\usepackage{m-pictex, m-ch-en}
\usepackage {amsmath}
\usepackage[squaren, Gray, cdot]{SIunits}
\usepackage{color}

\begin{document}


\title{Origin of molecular oxygen in Comet 67P/Churyumov-Gerasimenko}


\author{O. Mousis\altaffilmark{1}, T. Ronnet\altaffilmark{1}, B. Brugger\altaffilmark{1}, O. Ozgurel$^{2}$, F. Pauzat\altaffilmark{2}, Y. Ellinger\altaffilmark{2}, R. Maggiolo\altaffilmark{3}, P. Wurz\altaffilmark{4}, P. Vernazza\altaffilmark{1}, J. I. Lunine\altaffilmark{5}, A. Luspay-Kuti\altaffilmark{6}, K. E. Mandt\altaffilmark{6}, K. Altwegg\altaffilmark{4}, A. Bieler\altaffilmark{4}, A. Markovits$^{2}$, and M. Rubin\altaffilmark{4}}


\altaffiltext{1}{Aix Marseille Universit{\'e}, CNRS, LAM (Laboratoire d'Astrophysique de Marseille) UMR 7326, 13388, Marseille, France {\tt olivier.mousis@lam.fr}}
\altaffiltext{2}{Laboratoire de Chimie Th\'eorique, Sorbonne Universit\'es, UPMC Univ. Paris 06, CNRS UMR 7616, F-75252 Paris CEDEX 05, France}
\altaffiltext{3}{Royal Institute for Space Aeronomy, 3 avenue circulaire, Brussels, Belgium}
\altaffiltext{4}{Physikalisches Institut, University of Bern, Sidlerstrasse 5, CH-3012 Bern, Switzerland}
\altaffiltext{5}{Department of Astronomy and Carl Sagan Institute, Space Sciences Building Cornell University,  Ithaca, NY 14853, USA}
\altaffiltext{6}{Department of Space Research, Southwest Research Institute, 6220 Culebra Rd., San Antonio, TX 78228, USA}


\begin{abstract}

Molecular oxygen has been detected in the coma of comet 67P/Churyumov-Gerasimenko with abundances in the 1--10\% range by the ROSINA-DFMS instrument on board the Rosetta spacecraft. Here we find that the radiolysis of icy grains in low-density environments such as the presolar cloud may induce the production of large amounts of molecular oxygen. We also show that molecular oxygen can be efficiently trapped in clathrates formed in the protosolar nebula, and that its incorporation as crystalline ice is highly implausible because this would imply much larger abundances of Ar and N$_2$ than those observed in the coma. Assuming that radiolysis has been the only O$_2$ production mechanism at work, we conclude that the formation of comet 67P/Churyumov-Gerasimenko is possible in a dense and early protosolar nebula in the framework of two extreme scenarios: (1) agglomeration from pristine amorphous icy grains/particles formed in ISM and (2) agglomeration from clathrates that formed during the disk's cooling. The former scenario is found consistent with the strong correlation between O$_2$ and H$_2$O observed in 67P/C-G's coma while the latter scenario requires that clathrates formed from ISM icy grains that crystallized when entering the protosolar nebula.

\end{abstract} 


\section{Introduction}

The Rosetta Orbiter Spectrometer for Ion and Neutral Analysis (ROSINA) Double Focusing Mass Spectrometer (DFMS) on board the Rosetta spacecraft (Balsiger et al. 2007) enabled the detection of O$_2$ in the coma of comet 67P/Churyumov-Gerasimenko (67P/C-G) with local abundances in the 1--10\% range and a mean value of 3.80 $\pm 0.85\%$ (Bieler et al. 2015). A subsequent reinvestigation of the 1P/Halley data from the Giotto Neutral Mass Spectrometer (NMS) also indicates that the coma of comet 1P/Halley should contain O$_2$ with an abundance of 3.7 $\pm$ 1.7\% with respect to water, suggesting that this molecule may be a rather common parent species in comets (Rubin et al. 2015a).

To investigate the origin of O$_2$ in 67P/C-G, Bieler et al. (2015) considered the possibility of O$_2$ production via the radiolysis of water ice incorporated within the nucleus. Based on 67P/C-G's known orbital history, they estimated that any O$_2$ produced during the residence time of 67P/C-G in the Kuiper Belt was quickly lost during the first pass or two around the Sun. The authors further found that radiolysis on closer orbit to the Sun would most likely only affect the top few micrometres of the nucleus' active surface. In this case the O$_2$/H$_2$O ratio produced in these conditions would decrease with depth. Because they did not observe any variation of the O$_2$/H$_2$O ratio during the sampling period, Bieler et al. (2015) ruled out the hypothesis of O$_2$ production via the radiolysis and determined that O$_2$ must have been incorporated into 67P/C-G at the time of its formation in the protosolar nebula (PSN). 

In order to explain how O$_2$ could have been incorporated into the ices of 67P/C-G, we investigate here the radiolysis hypothesis at epochs prior to the formation of comets, when icy grains were the dominant solid phase of the outer PSN. Furthermore, we examine the different trapping scenarios of O$_2$ that could explain its presence. Because some recent works suggest that this comet may have been accreted from a mixture of clathrates and pure crystalline ices formed in the PSN (Mousis et al. 2016; Luspay-Kuti et al. 2016), we study the propensity for O$_2$ trapping in clathrates, and also evaluate if its condensation as pure crystalline ice is consistent with the comet's inferred composition. Among all these investigated mechanisms, we find that clathration of O$_2$ is efficient in the PSN and that radiolysis can explain the formation of O$_2$ and its stabilization in icy grains. However, to produce enough O$_2$ molecules, the radiolysis of icy grains must have happened in a low density environment such as the presolar cloud.
 
\section{O$_2$ formation via radiolysis}

We first investigate the possibility of radiolytic production of O$_2$ in icy grains present in the outer PSN prior to their agglomeration by 67P/C-G. The energy available for radiolysis is provided by the galactic Cosmic Ray Flux (CRF) impacting icy grains. {In the following, since galactic CR can penetrate into water ice down to depths of a few tens of meters (Cooper et al. 1998), we only consider icy grains with sizes below this limit, implying that no H$_2$O ice can be out of reach of radiolysis.} In our calculations, we use the energy range and CRF distribution from Yeghikyan (2011) and Cooper~et~al. (2003), respectively. The CRF energy dose absorbed by icy grains located at 30~AU from the sun is within the $\sim$(5--60)$\times$10$^{16}$~eV~kg$^{-1}$~yr$^{-1}$ range, depending on the disk's surface density (between 10 and 10$^3$~g~cm$^{-2}$; see Hersant~et~al. 2001).

O$_2$ is produced by radiolysis of water ice through the chemical reaction 2~H$_2$O~$\longrightarrow$~2~H$_2$~+~O$_2$, with an amount of energy needed to alter one H$_2$O molecule $W_r$~=~235~eV (Johnson 1991). H$_2$ is then rapidly lost from the water ice due to its fast diffusion. Further reactions with O$_2$ are precluded because the diffusion of these molecules is slowed down by the disk's low temperatures (Johnson 1990). We have thus assumed that all the energy absorbed by water ice is used to form O$_2$. To reach the molecular ratio O$_2$/H$_2$O (1--10\%) measured by Bieler~et~al. (2015) in 67P/C-G, cosmic rays must alter twice as many H$_2$O molecules in icy grains. The time $\tau$ needed to reach this ratio is then given by:

\begin{equation}
\tau = \dfrac{W_r \cdot N_A}{E_{CR} \cdot M_{H_2O}} \times f_{H_2O}
\end{equation}

\noindent where $N_A$ (mol$^{-1}$) is the Avogadro constant, $M_{H_2O}$~(kg mol$^{-1}$) is the molar mass of water, $E_{CR}$ (eV~kg$^{-1}$~yr$^{-1}$) is the CRF energy dose received by water ice and $f_{H_2O}$ is the fraction of altered H$_2$O molecules, which corresponds to two times the fraction of O$_2$ produced.

\begin{figure}[h]
\begin{center}
\resizebox{\hsize}{!}{\includegraphics[angle=0]{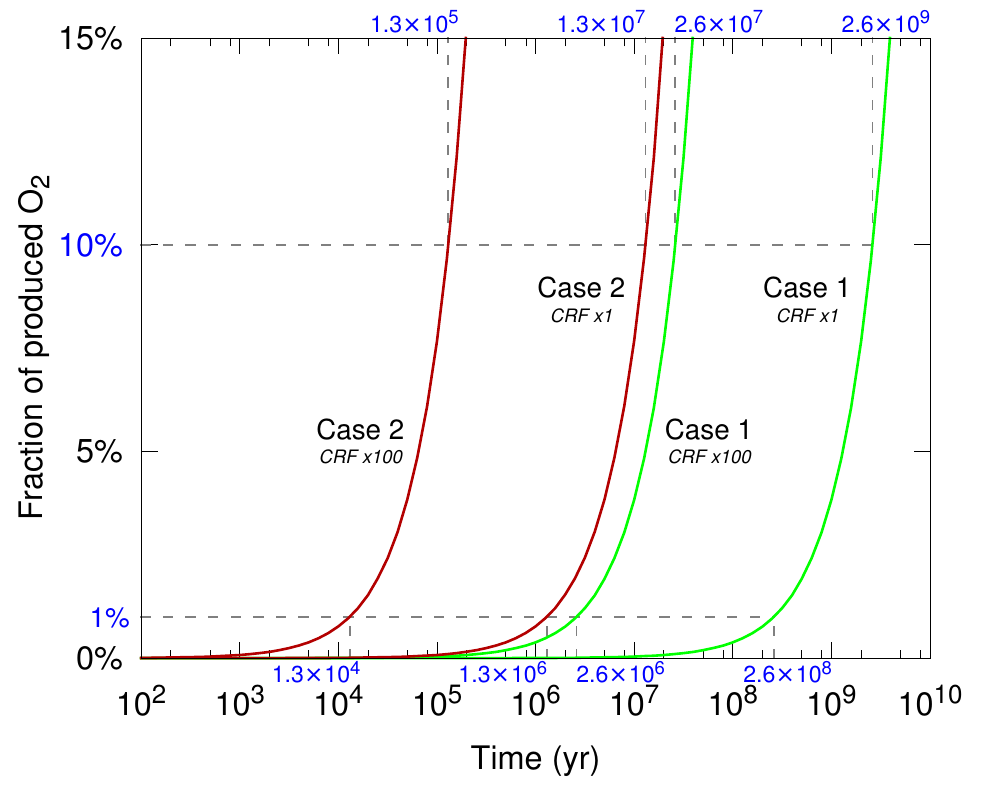}}
\caption{Time evolution of the fraction of O$_2$ molecules produced by cosmic rays in an icy grain. Case~1 (green curves) considers the irradiation of an icy grain placed at 30~AU in the PSN. Case~2 (red curves) considers the irradiation of an icy grain located in a low density environment ($\sim$10$^{-3}$~g~cm$^{-2}$). Two CRF values are explored in each case, namely 1 and 100 times the nominal CRF value (see text).}
\label{fig:cosmic_rays}
\end{center}
\end{figure}

Figure~\ref{fig:cosmic_rays} shows the results of our calculations. An O$_2$ fraction in the 1--10\% range is reached in $\sim$0.25--2.5~Gyr at the aforementioned nominal CRF value (Case~1). These extremely long time periods are incompatible with the lifetime of icy grains in the PSN {(a few 10$^4$ yr; Weidenschilling and Cuzzi 1993). If icy grains have grown to sizes larger than tens of meters in the PSN, then the deepest layers should remain unaltered. In this case, even longer timescales would be needed for O$_2$ formation.} However, the CRF may have undergone significant enhancements throughout the history of the solar system, by a factor $\sim$3 during its passages through the Milky Way's spiral arms (a few tens of Myr every 400--500~Myr; Effenberger~et~al. 2012; Werner~et~al. 2015; Alexeev 2016), or even by a factor of $\sim$100 during a few kyr because of a close supernova explosion ($<$30~pc; Fields~\&~Ellis 1999). Such enhancements can decrease the time needed to form O$_2$ by up to a factor 100, which is still too long for our consideration.

We also consider the possibility of an icy grain receiving the maximum CRF energy dose estimated by Yeghikyan (2011), namely $\sim$1.20$\times10^{20}$~eV~kg$^{-1}$~yr$^{-1}$. This value leads to a time $\tau$ in the $\sim$1--10~Myr range (see case~2 of Fig.~\ref{fig:cosmic_rays}), or $\sim$10--100~kyr with a CRF enhanced by a factor 100. However, such a high value of $E_{CR}$ corresponds to a surface density of 10$^{-3}$~g~cm$^{-2}$, which can only be reached in molecular clouds. In such environments, the column densities would be low enough to form 1--10\% of O$_2$ in the icy grains even on very short timescales. Therefore, to incorporate significant amounts of O$_2$ produced via radiolysis of icy grains, cometary grains must have formed in the presolar cloud prior to disk formation.

\section{O$_2$ stability in water ice}

\begin{figure}[h]
\begin{center}
\resizebox{.8\hsize}{!}{\includegraphics[angle=0]{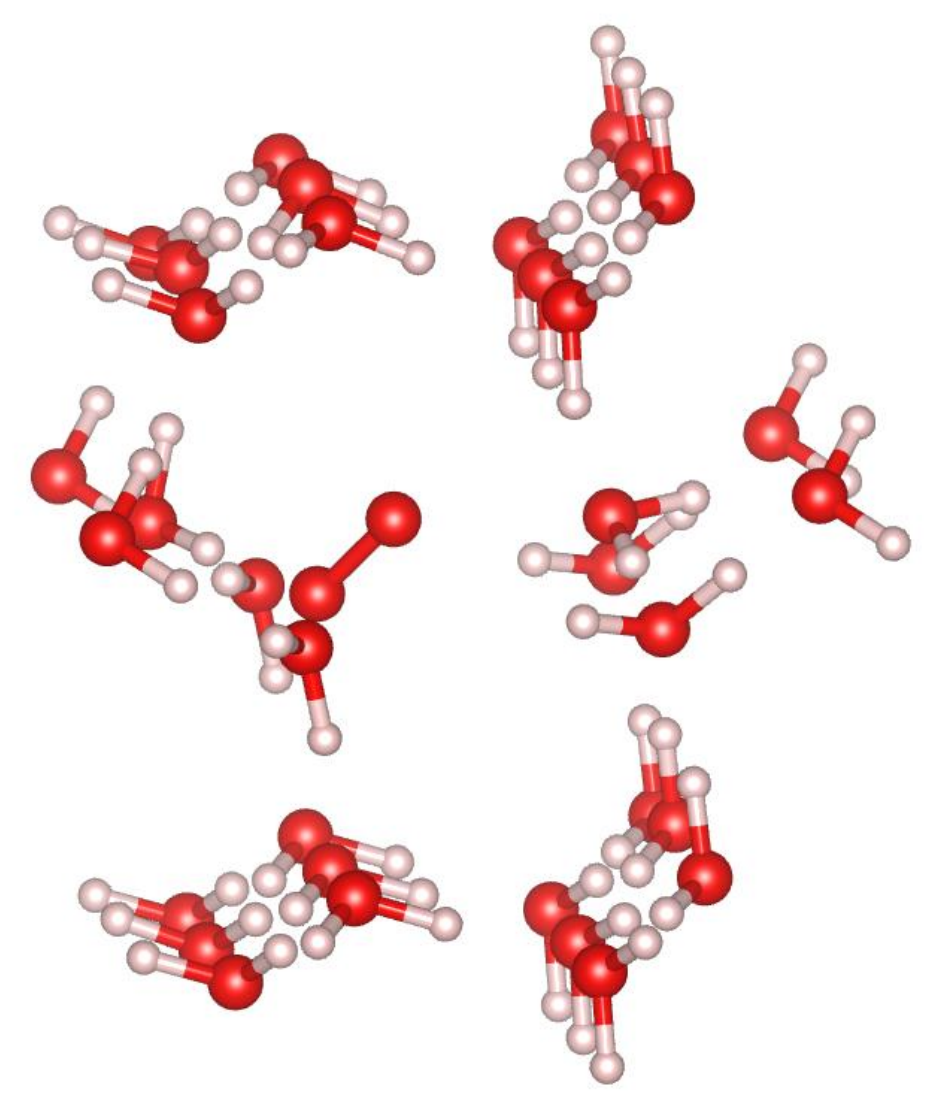}}
\caption{Side view of O$_2$ embedded in a cavity  inside compact amorphous ice. The cavity corresponds to a void of  3 H$_2$O molecules from an hexagonal apolar lattice.}
\label{holes}
\end{center}
\end{figure}

An important question is whether O$_2$ molecules produced via radiolysis of ice grains can remain stabilized within the water icy matrix of 67P/C-G. {The stabilization energy is defined as the difference between the energy of the system of O$_2$ interacting with the ice and the sum of the energies of the pure ice and O$_2$ at infinite separation.} To investigate this problem, a sampling of the representative structures of O$_2$ in solid water ice has been obtained using a strategy based on {first principle periodic density functional theory quantum calculations}, that has been proven to be appropriate for modeling bulk and surface ice structures (Lattelais et al. 2011; 2015; Ellinger et al. 2015). Among the different  forms, we considered the apolar variety of hexagonal ice {\it Ih} {because these structures have a balanced distribution of alternating hydrogen and oxygen avoiding computational artifacts for surface optimizations and at the same time reproduce the bulk properties (Casassa et al. 2005).} How O$_2$ behaves as a function of the number of H$_2$O molecules removed is illustrative of the storage capability of the ice as a function of porosity. The results of our calculations, performed using the Vienna ab initio simulation package (VASP) (Kresse \& Hafner 1993; 1994), are presented below:

\begin{enumerate}
\item Starting with no H$_2$O removed, i.e., the pure cristalline ice, we found no stabilisation for the inclusion of O$_2$ in the hexagonal lattice. It is in fact an endothermic process.

\item With one H$_2$O removed, and replaced by one O$_2$, we have a substitution structure whose stabilisation, in the order of 10$^{-3}$ eV,  is meaningless.

\item With 2, 3, and 4 adjacent H$_2$O molecules removed from the hexagonal lattice we obtained well defined cavities that, after reconstruction, show  different shapes according to the positions of the entities removed. The stabilization energies were found to be on the order of 0.2--0.3 eV, going to  0.4--0.5 eV for an embedded O$_2$ dimer. A typical structure of embedding is illustrated in Fig. \ref{holes} where  O$_2$ is stabilized with an energy of $\sim$0.23 eV. This energy is of the order of that of a water dimer which means that the presence of O$_2$ should not perturb the ice structure until it is ejected into the coma via sublimation with the surrounding  H$_2$O molecules.
\end{enumerate}

It should be stressed that the formation of one O$_2$ requires at least the destruction of two H$_2$O. The present simulation is fully consistent with the aforementioned radiolysis hypothesis, where the irradiation process is at the origin of both the formation of O$_2$ and the development of the cavity in which it remains sequestered. Similar results are obtained in the case of O$_2$ stabilization in amorphous ice.

\section{O$_2$ clathration in the PSN}
\label{cla}

One possible source of O$_2$ in the nucleus of 67P/C-G is the trapping of O$_2$ in clathrates that formed in the PSN prior to having been agglomerated by the comet as it formed. This is supported by recent works showing that the Ar/CO and N$_2$/CO ratios and the time variation of other volatile species measured in 67P/C-G's coma are found to be consistent with the presence of clathrates in its nucleus (Mousis et al. 2016; Luspay-Kuti et al. 2016). To investigate the amount of O$_2$ that could have been trapped in clathrates and now be present in 67P/C-G, we use the same statistical thermodynamic model as the one described in Mousis et al. (2010, 2016), which is used to estimate the composition of these crystalline structures formed in the PSN. To evaluate the trapping efficiency of O$_2$, we consider a gas constituted of O$_2$ and CO. After H$_2$O, CO is one of the dominant gases found in 67P/C-G (Le Roy et al. 2015) and in most of comets (Bockel\'ee-Morvan et al. 2004). The Kihara parameters for the molecule-water interactions employed in our calculations are derived from Mohammadi et al. (2003) for O$_2$ and from Mohammadi et al. (2005) for CO. These represent the most recent sets of data found in the literature for the two species. We refer the reader to the model description provided in Mousis et al. (2010) for further details. 

\begin{figure}[h]
\begin{center}
\resizebox{\hsize}{!}{\includegraphics[angle=0]{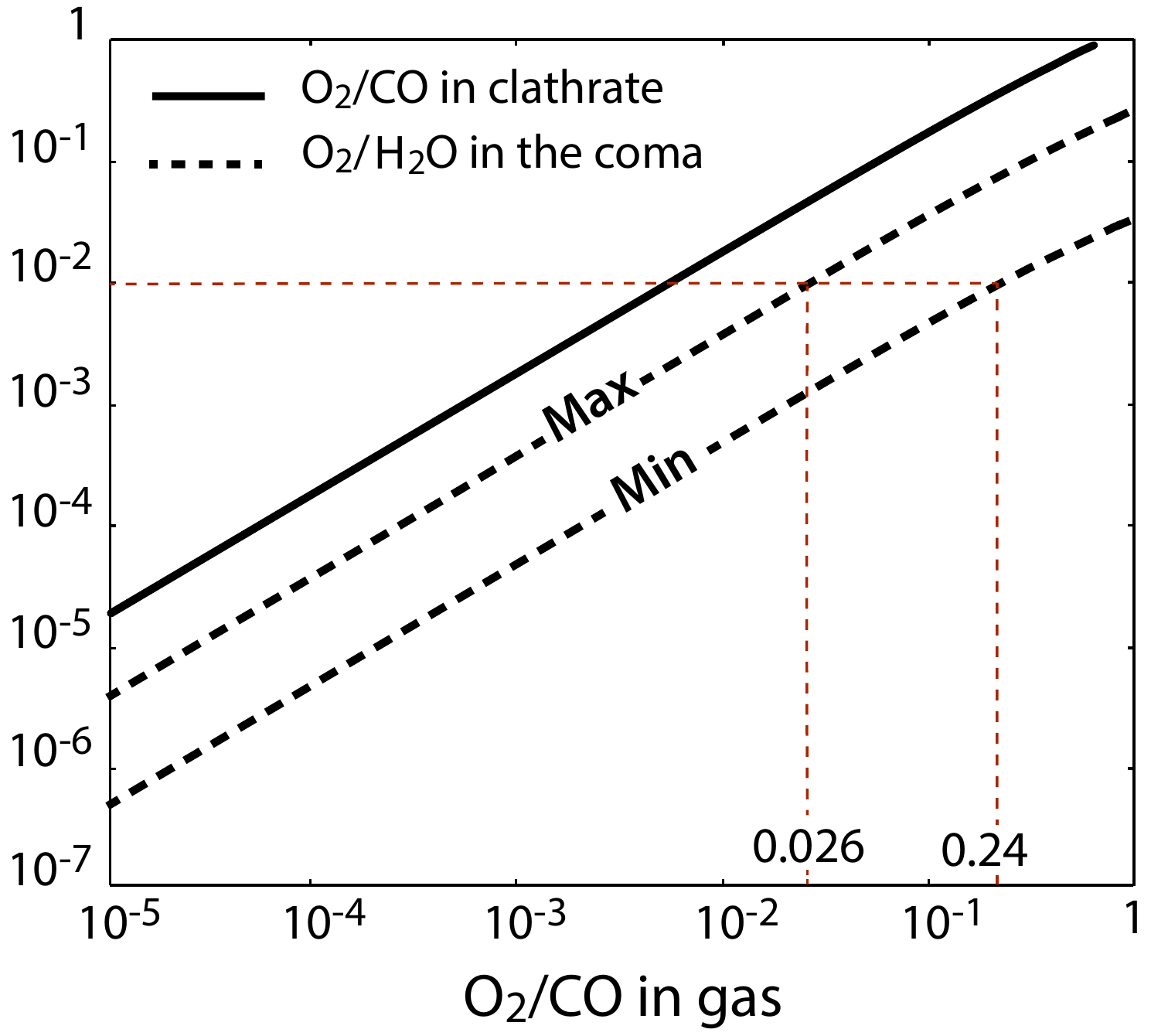}}
\caption{O$_2$/CO ratio in clathrates formed at 45 K and the corresponding O$_2$/H$_2$O ratio in the coma, as a function of the coexisting O$_2$/CO ratio in the PSN gas phase. The ``Min'' and ``Max'' labels correspond to calculations of the O$_2$/H$_2$O ratio in 67P/C-G's coma, assuming that the CO/H$_2$O abundance is between 2.7 and 21\% (see text). The vertical red dashed lines represent the O$_2$/CO ratio in the PSN gas phase needed to form clathrates giving 1\% O$_2$ relative to H$_2$O in the coma.}
\label{clat}
\end{center}
\end{figure}

When clathrates destabilize in the nucleus, the trapped volatiles are released prior to water sublimation, implying that the water vapor measured at the time of the O$_2$ sampling by ROSINA should be derived from the vaporization of crystalline ice layers located closer to the surface. Hence, the O$_2$ depletion is better quantified by comparing the O$_2$/CO ratio in clathrates and the coma value since these two species are expected to be released simultaneously from destabilized clathrates. Figure \ref{clat} represents the value of the O$_2$/CO ratio in structure I clathrates\footnote{Both O$_2$ and CO molecules are expected to form this structure (Mohammadi et al. 2003, 2005).} as a function of the O$_2$/CO ratio in the coexisting gas phase at a chosen disk's temperature of $\sim$45 K. This value is within the temperature range needed for clathrates to form in the PSN from a gaseous mixture of protosolar composition that reproduces the Ar/CO and N$_2$/CO ratios measured in 67P/C-G's coma (Mousis et al. 2016). We find that, whatever the O$_2$/CO ratio considered for the initial PSN gas phase, it is enriched by a factor of $\sim$1.4--1.8 in the formed clathrate. Figure \ref{clat} also shows that the O$_2$/CO ratio must be in the 0.026--0.24 range in the PSN gas phase for the clathrate trapping mechanism to agree with the measured range of O$_2$/H$_2$O in the coma ($\sim$1\%), assuming that all cavities are filled by guest molecules and that the CO/H$_2$O abundance ratio in the coma corresponds to the sampled value ($\sim$2.7--21\%; Le Roy et al. 2015). {This range of O$_2$/CO ratios is consistent with values obtained at distances beyond $\sim$5 AU in a T Tauri disk (Walsh et al. 2015).} Therefore, our calculations show that the clathration of O$_2$ in the PSN is a realistic mechanism to account for the O$_2$/H$_2$O ratio observed by ROSINA in 67P/C-G's coma.

\section{O$_2$ condensation in the PSN}
\label{cond}

\begin{figure}[h]
\begin{center}
\resizebox{\hsize}{!}{\includegraphics[angle=0]{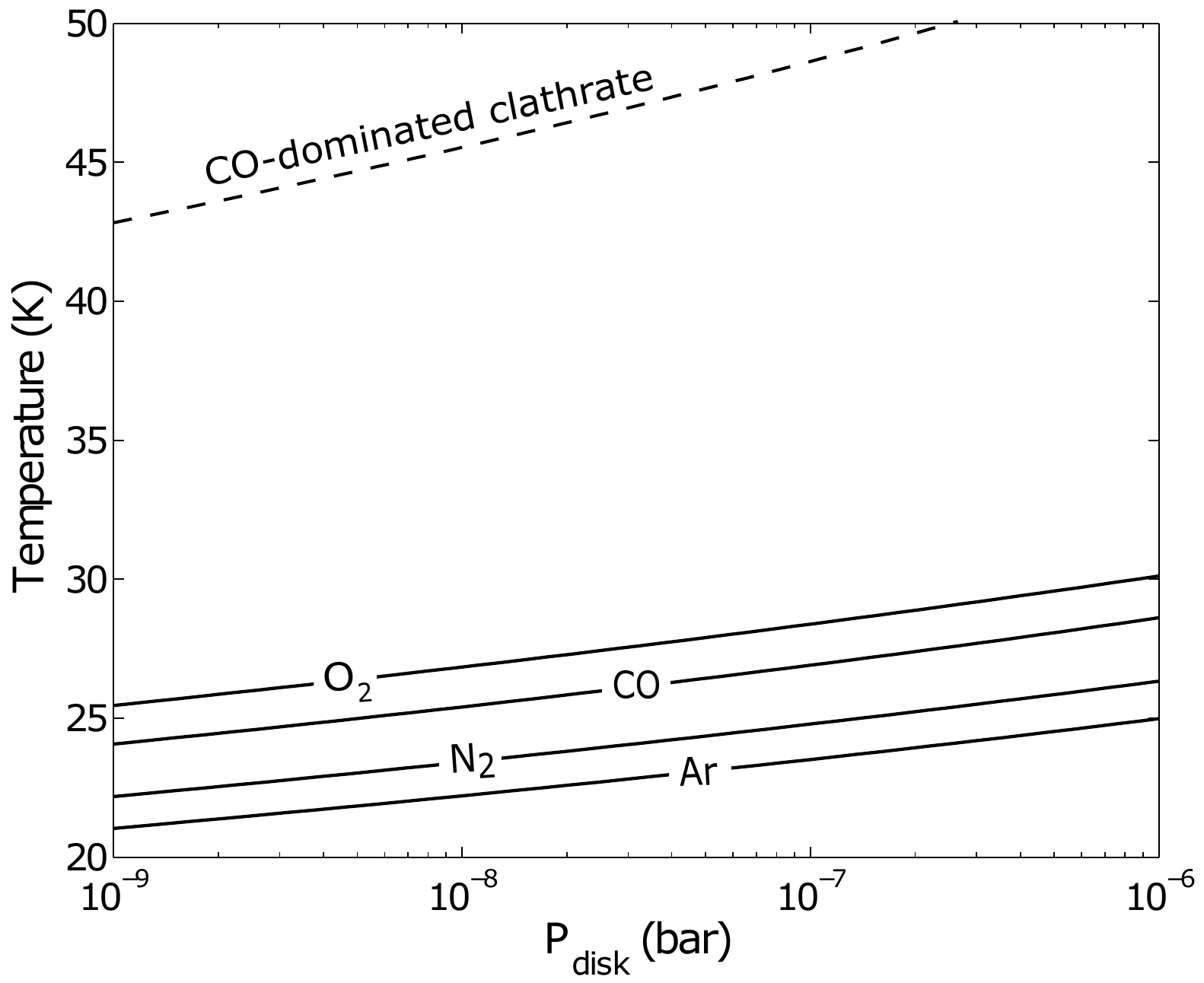}}
\caption{Solid lines: equilibrium curves of O$_2$, CO, N$_2$ and Ar pure crystalline ices as a function of total disk pressure. Dashed line: equilibrium curve of the CO-dominated clathrate as a function of total disk pressure (see text).}
\label{curv}
\end{center}
\end{figure}

An alternative possibility for the observed presence of O$_2$ in the coma of 67P/C-G is that the O$_2$ could have been agglomerated as pure crystalline ice by the nucleus forming at cooler PSN temperatures than those required for clathration. To investigate this scenario, we calculated the temperature dependence of the equilibrium curves of O$_2$, CO, N$_2$ and Ar pure crystalline ices via the use of the polynomial relations reported by Fray \& Schmidt (2009). To derive the partial pressures for each gas, we assumed that O, C, N and Ar exist in protosolar abundances in the PSN (Lodders et al. 2009), with all C and all N in the forms of CO and N$_2$, respectively. The partial pressure of O$_2$ is derived from the O$_2$/CO gas phase ratio ($\sim$33\%) predicted beyond the snowline of a T Tauri disk via an extensive chemical model (Walsh et al. 2015). The equilibrium curves of O$_2$, CO, N$_2$ and Ar pure crystalline ices are represented along with the equilibrium curve of the CO--N$_2$--Ar multiple guest clathrate proposed by Mousis et al. (2016) to explain 67P/C-G's composition, as a function of the total PSN pressure in Fig. \ref{curv}. Because the CO--N$_2$--Ar multiple guest clathrate is by far dominated by CO (see Fig. 1 of Mousis et al. 2016), we assume that its partial pressure is the same as for CO crystalline ice. The equilibrium curve of the clathrate is taken from Lectez et al. (2015).

From the examination of the condensation sequence presented in Fig. \ref{curv}, we find that the hypothesis of O$_2$ agglomeration as pure crystalline ice is inconsistent with 67P/C-G's current composition. The fact that Ar/CO and N$_2$/CO ratios are found to be significantly depleted by factors of $\sim$90 and 10 in 67P/C-G's coma, respectively, compared to the protosolar values (Rubin et al. 2015b; Balsiger et al. 2015; Mousis et al. 2016), implies that Ar and N$_2$ cannot form substantial amounts of pure crystalline ices at the formation location of the comet in the PSN (Mousis et al. 2016). Instead, it has been proposed that these volatiles were mostly trapped in CO-dominated clathrates (Mousis et al. 2016). Under these circumstances, because the equilibrium curve of O$_2$ ice is in the vicinity of those of Ar and N$_2$ ices, the incorporation of O$_2$ in this form would require the trapping of larger amounts of Ar and N$_2$, incidentally leading to quasi protosolar Ar/CO and N$_2$/CO ratios. This does not agree with the depleted ratios observed in 67P/C-G.

\section{Conclusions}

In this study, we have investigated several scenarios that may explain the presence of molecular oxygen in the nucleus of 67P/C-G. Our results are the following:

\begin{itemize}

\item Even with a strong CRF enhancement due to the presence of a nearby supernova, we find that the radiolysis of icy grains is not fast enough in the PSN to create amounts of O$_2$ comparable with those observed in 67P/C-G. Instead, icy grains must be placed in low-density environments such as molecular clouds to allow radiolysis to work efficiently. The irradiation process also favors the stabilization of O$_2$ molecules in the icy matrix via the development of cavities and is compatible with both amorphous and crystalline ice structures.

\item O$_2$ can be efficiently trapped in clathrates formed in the PSN. The O$_2$/CO ratio in the clathrate phase is up to $\sim$2 times the O$_2$/CO ratio in the coexisting PSN gas phase.

\item The incorporation of O$_2$ as pure crystalline ice is unlikely in 67P/C-G because the condensation of this species in the PSN would imply much larger abundances of Ar and N$_2$ than those observed in the coma. 

\end{itemize}

Based on these results, and assuming that radiolysis has been the only mechanism for producing O$_2$, we find that the formation of 67P/C-G is possible in a dense and early PSN in the framework of two extreme scenarios: (1) agglomeration from pristine amorphous icy grains/particles formed in the ISM and (2) agglomeration from multiple guest clathrates including O$_2$ that formed during the cooling of the disk subsequent to the vaporization of the amorphous icy grains entering the PSN. However, scenario 1 was found inconsistent with ROSINA pre-perihelion observations of volatile abundances in the coma. In contrast, Mousis et al. (2016) and Luspay-Kuti et al. (2016) have shown that scenario 2 could match these data if 67P/C-G agglomerated from a mixture of clathrates and crystalline ices that condensed in the PSN. Also, scenario 2 is compatible with a possible chemical production of O$_2$ in the PSN gas phase (Walsh et al. 2015). In this picture, whatever the considered source, i.e. radiolysis of ISM grains or/and PSN gas phase chemistry, O$_2$ is efficiently entrapped in clathrates prior to their agglomeration by 67P/C-G. 

On the other hand, with the incorporation of O$_2$ in the cavities created by CRF in the icy matrix, scenario 1 naturally provides an explanation for the strong correlation found between the O$_2$ and H$_2$O production rates observed in 67P/C-G's coma (Bieler et al. 2015). If this scenario is correct, this would make implausible the accretion of 67P/C-G from clathrates and crystalline ices originating from the PSN. Meanwhile, a way to reconcile scenario 2 with the strong O$_2$-H$_2$O correlation would be to assume that the icy grains initially formed as in scenario 1. These icy grains/particles would have then subsequently experienced an amorphous-to-crystalline phase transition in the 130--150 K temperature range when entering the disk (Kouchi et al. 1994; Maldoni et al. 2003; Ciesla 2014). In this alternative scenario, all volatiles initially adsorbed by ISM amorphous ice would be released in the PSN gas phase during phase transition. With the cooling of the disk, these volatiles would have been later trapped in the clathrates formed with the crystallized icy grains. The case of O$_2$ is unique because, due to its formation process, this molecule is inserted into the icy matrix. In spite of the phase transition, O$_2$ would remain stable within the icy matrix because the strength of the interaction between O$_2$ and the surrounding H$_2$O molecules is expected not to decrease (eventually increase) upon crystallization. In this scenario, CO, Ar, N$_2$ would be trapped in clathrates with O$_2$ remaining embedded in water, in a way consistent with the observed correlation.

To conclude, further post-perihelion ROSINA data, in particular the precise measurements of the relative abundances of the different volatiles as a function of geography and time, are needed to disentangle between the existing formation scenarios. It is also possible that only the in situ sampling of a nucleus by a future lander will provide a definitive answer to the question of the formation conditions of 67P/C-G and other Jupiter Family Comets in the PSN.

\acknowledgements
O.M. acknowledges support from CNES. This work has been partly carried out thanks to the support of the A*MIDEX project (n\textsuperscript{o} ANR-11-IDEX-0001-02) funded by the ``Investissements d'Avenir'' French Government program, managed by the French National Research Agency (ANR). This work also benefited from the support of CNRS-INSU national program
for  planetology (PNP). R.M. was supported by the Belgian Science Policy Office through the Solar-Terrestrial Centre of Excellence and by PRODEX/ROSETTA/ROSINA PEA 4000107705. J.I.L. acknowledges support from JWST. K.E.M. acknowledges support from JPL Subcontract 1345493.



\end{document}